\documentclass[epj]{svjour}
\usepackage{graphicx}
\usepackage{amssymb}

\begin{document}

\title{Decay of entanglement in coupled, driven systems with bipartite decoherence}

\author{J. Li and G.~S.~Paraoanu}

\institute{Low Temperature Laboratory, Helsinki University of Technology,
P.O. Box 5100, FIN-02015 TKK, Finland}

\date{}

\abstract{
We analyze a system of two qubits embedded in two different environments. The qubits are coupled to each other and driven on-resonance by two external classical sources.
In the secular limit, we obtain exact analytical results for the evolution of the system for several classes of two-qubit mixed initial states. For Werner states we show that the decay of entanglement does not depend on coupling.  For other initial states with \lq\lq{\sf X}\rq\rq -type density matrices we find that the sudden death time displays a rich dependence on the coupling energy and state parameters due to the existence of processes of delayed sudden birth of entanglement.
}


\maketitle

\section{Introduction}

In the theory of open quantum systems, when a two-level system (a qubit)
is placed in a dissipative environment (such as a bath of harmonic oscillators), a number of decoherence effects are known to occur and the mathematical formalism describing them has been worked out in the last few decades \cite{book}.
A very interesting situation occurs when one considers two qubits and their entangling properties: if the qubits are coupled to two different (uncorrelated) baths, qualitatively new effects, known as {\it sudden death of entanglement} phenomena, have been found to occur \cite{Yu1,Yu3,Yu2}. These effects, already confirmed experimentally \cite{almeida}, include the non-exponential decay of concurrence \cite{Yu1} and the nonadditivity of decay rates \cite{Yu3}. It has been shown that these sudden death effects becomes more pronounced at finite temperature \cite{finitetemp}
and in the presence of external driving \cite{driven}. Such systems have attracted considerable interest amongst various research communities, due to the fact that the entangling effects predicted are fundamental and generic  - in the sense that they do not depend on a specific experimental realization.

Much of the nonituitive character of the sudden death of entanglement comes from the fact that the system is bipartite and, although it is subject only to local
 interactions (acting separately on each of the two subsystems), one property (entanglement) which characterizes the whole system,
has a dynamics which cannot be understood qualitatively as just the addition of local effects. There are also practical consequences to the sudden death effect:
for the research effort into building a quantum computer, it implies that it could be more difficult to maintain entanglement when the qubits interact with different reservoirs and are locally driven by single-qubit quantum gates. In this context, also qubit-qubit interactions must be considered: in quantum networks such as those realized nowadays with superconducting quantum circuits these interactions can appear while performing two-qubit gates or simply by spurious cross-coupling.  This qubit-qubit coupling  is established either directly, through mutual inductances and capacitances \cite{direct}, or as second-order effects, due to the creation of virtual excitations in the common environment or in the quantum buses needed in a
typical architecture of a quantum processor \cite{indirect}. Finally, in the presence of driving and of two different reservoirs, the qubit-qubit interaction leads to another entanglement effect called steady-state generation of entanglement, a phenomenon which has attracted considerable attention \cite{steadystate}. In contrast, transient entanglement effects that occur as shorter timescales in such systems have not yet been studied systematically. Much like in the case of steady-state entanglement, we will see that the interesting effects occur only when driving, coupling, and decoherence are present at the same time. However, we should mention that the steady-state generation of entanglement appears in a completely different region of the parameter space, in which the secular approximation is not valid: for the parameter values considered here, the steady state is not entangled.

In this paper, we investigate a system consisting of two qubits, dissipating in two different reservoirs, driven at resonance, and coupled by a dipole-dipole interaction.
We study the entanglement dynamics of initial states of \lq\lq{\sf X}\rq\rq -type, that is, $4\times 4$ density matrices in which the nonzero elements have the shape of the letter "X".
As we will see, under unitary evolution and in the secular approximation these states are mapped also into X-type states and finding analytical solutions is possible. For the states investigated, the dynamics occurs at timescales below the inverse of the longitudinal decoherence rate -- the states become separable after a time
of sudden death $t_{ESD}$.  Bipartite decoherence is truly a requirement: for a single reservoir, the decay would be exponential, which  would make any diagram of $t_{ESD}$ featureless. As it turns out, the three processes present in the problem - driving, coupling, and decoherence - combine into a non-trivial time evolution in the interval $t\in [0, t_{ESD}]$ resulting in an unexpectedly rich dependence of $t_{ESD}$ on the initial state and on the coupling strength.

We consider two qubits of energy separation (Larmor frequency) $\omega_j^L$, and driven by two classical radiation fields of frequencies $\omega_j$ which are coupled to the $\sigma^x$-Pauli matrices of the qubits
with dipole-field strengths (Rabi frequencies) $\Omega_j$ .
The qubits are also dipole-dipole coupled to each other, a generic interaction which appears as such in many experimentally-relevant contexts \cite{direct,indirect}. In the Schr\"odinger picture we write for this part of the Hamiltonian
\begin{equation}
H_{coup} = \omega^{xx}\sigma_1^x\sigma_2^x + \omega^{yy}\sigma_1^y\sigma_2^y ,\label{eq_coupling}
\end{equation}
where $\omega^{xx}$ and $\omega^{yy}$ denotes the X-X and Y-Y coupling strengths, respectively.
The total Hamiltonian of the system is therefore
\begin{equation}
H_{sys} = \sum_{j=1,2} \frac{\omega_j^L}{2}\sigma_j^z + \sum_{j = 1,2} \Omega_j\cos(\omega_j t)\sigma_j^x +  H_{coup}
\end{equation}

For the environment, we consider a longitudinal (amplitude damping) reservoir at zero temperature.
Under the usual assumptions of Markovian evolution with respect to the environment degrees of freedom \cite{book},
and at zero temperature,
the dynamical evolution of the qubits' density operator (in the Schr\"odinger picture)
$\rho_S$ is governed by the Born-Markov master equation
\begin{eqnarray}
\dot{\rho}_S = -i [H_{sys}, \rho_S]+  {\cal L}[\rho_S] , \label{eq_original_master_equation}
\end{eqnarray}
where the resulting Liouvillean superoperator $\cal{L}$ is the sum of the longitudinal dampings of
the two qubits with energy relaxation rates $\Gamma_j$, respectively,
\begin{eqnarray}
{\cal L}[\rho_S] = \sum_{j=1,2} \frac{\Gamma_j}{2} \left( 2\sigma_j^- \rho_S \sigma_j^+ - \sigma_j^+ \sigma_j^-
\rho_S - \rho_S \sigma_j^+
\sigma_j^- \right)
\end{eqnarray}

We note here that we have implicitly used the typical assumption (largely satisfied for many experimental systems) that the energy relaxation rates $\Gamma_j$ have a mild, negligible variation with frequency over a bandwidth of the order of $\Omega_j$ around the Larmor frequencies of the qubits (or, leading to the same conclusion, one assumes that driving is weak with respect to the Larmor frequencies). We then move to a rotating frame using the transformation
$R = \exp[i(\omega_1\sigma_1^z + \omega_2\sigma_2^z)t / 2]$ and the identities $R^\dag \sigma_{j}^{\pm}R = \sigma_{j}^{\pm}
\exp (\mp i\omega_{j}t)$
with $j=1,2$, and also employing the rotating-wave approximation to eliminate the relatively faster counter-rotating terms,
\begin{equation}
\dot{\rho}_{rf} = -i [H_{rf}, \rho_{rf}] + {\cal L}[\rho_{rf}] , \label{eq_master_equation_in_rf}
\end{equation}
with $\rho_{rf} = R\rho_S S^\dag$, and
\begin{eqnarray}
H_{rf} &=& R H_{sys} R^{\dag} + i\frac{\partial R}{\partial t}R^{\dag} \\
&\approx &\sum_{j=1,2}\left( \frac{\delta_j}{2}\sigma_j^z + \frac{\Omega_j}{2}\sigma_j^x \right) +
\frac{\omega_{c}}{2}\left( \sigma_1^x\sigma_2^x + \sigma_1^y\sigma_2^y\right). \label{eq_h_rf}
\end{eqnarray}
Here $\delta_j = \omega_j^L - \omega_j$ is the detuning of the qubit transition frequency from the driving frequency
and $\omega_{c}$ is the effective coupling frequency (in the rotating wave approximation) $\omega_{c}=\omega^{xx} + \omega^{yy}$.

 Eq. (\ref{eq_h_rf}) can be solved numerically and we will present such results below. As it turns out, the data from numerical simulations will display qualitative differences for different initial
 states, and just the numerical result would be unsatisfactory. It is in principle possible to solve analytically Eq. (\ref{eq_h_rf}): unfortunately the structure of even relatively general forms of the density matrices (such as the \lq\lq{\sf X}\rq\rq -form, see below) is not left invariant under evolution with Eq. (\ref{eq_h_rf}), and the results are very complicated.  To make progress, further simplifications are necessary. Below we will use the secular approximation, which is valid for most of the interesting experimental situations when typically the Rabi frequency is much larger than the energy relaxation rate, and we show that relatively easy to handle analytical results can be derived.

\section{Secular approximation for driven, coupled qubits}

In the rest of this paper we will analyze the case of on-resonance driving ($\delta_j = 0$) and
identical Rabi frequencies $\Omega_j=\Omega$. It is convenient to work in the interaction picture, defined by
\begin{equation}
\rho = \exp\left(\frac{i}{2}\sum_{j=1,2}\Omega_j\sigma_j^x t\right)\rho_{rf}\exp\left( -\frac{i}{2} \sum_{j=1,2}\Omega_j\sigma_j^x t\right) .
\end{equation}
Under this transformation, $\sigma_{j}^x$ is left invariant while $\sigma_z$ transforms as
\begin{equation}
\exp \left(\frac{i}{2}\Omega\sigma_j^x t\right) \sigma_{j}^y \exp \left(-\frac{i}{2}\Omega\sigma_j^x t\right) = \sigma_{y}\cos\Omega t - \sigma_j^{z}\sin \Omega t ,
\end{equation}
and we obtain a master equation with time-dependent dissipation,
\begin{eqnarray}
\dot{\rho} &=& -\frac{i\omega_{c}}{4}\left[ 2\sigma_1^x\sigma_2^x + \sigma_1^y\sigma_2^y + \sigma_1^z\sigma_2^z, \rho \right] \nonumber \\
& &
+ \frac{i}{4}\sum_{j=1,2}\Gamma_j \sin(\Omega t) \left( \sigma_j^+ \rho\sigma_j^z + \sigma_j^- \rho\sigma_j^z - \sigma_j^z \rho\sigma_j^+ \right.\nonumber \\
& & \left. - \sigma_j^z \rho\sigma_j^- + \sigma_j^+\rho + \rho\sigma_j^+ - \sigma_j^-\rho - \rho\sigma_j^- \right) \nonumber \\
&& + \frac{i}{8}\sum_{j=1,2}\Gamma_j \sin(2\Omega t) \left( \sigma_j^z \rho\sigma_j^- + \sigma_j^- \rho\sigma_j^z - \sigma_j^z \rho \sigma_j^+ - \sigma_j^+ \rho\sigma_j^z \right) \nonumber \\
&& + \frac{1}{16}\sum_{j=1,2} \Gamma_j [ 3 + 4\cos(\Omega t) + \cos(2\Omega t) ]\times \nonumber \\
&& \times \left( 2\sigma_j^-\rho\sigma_j^+ - \sigma_j^+\sigma_j^- \rho - 
\rho\sigma_j^+\sigma_j^- \right) \nonumber \\
&& + \frac{1}{16}\sum_{j=1,2} \Gamma_j [ 3 - 4\cos(\Omega t) + \cos(2\Omega t) ]\times \nonumber \\
&& \times \left( 2\sigma_j^+\rho\sigma_j^- - \sigma_j^-\sigma_j^+ \rho - 
\rho\sigma_j^-\sigma_j^+ \right) \nonumber \\
&& + \frac{1}{8}\sum_{j=1,2} \Gamma_j [ 1-\cos(2\Omega t) ] \left( \sigma_j^+ \rho\sigma_j^+ \right.\nonumber \\
& & \left. + \sigma_j^-\rho\sigma_j^- + \sigma_j^z
\rho\sigma_j^z - \rho \right) . \nonumber \label{eq_explicit_master_equation_in_ip}
\end{eqnarray}

We employ now the secular approximation, $\Omega\gg \Gamma_j$ \cite{Cohen}, which allows us to eliminate all the oscillating terms in the dissipative part $\Omega$ (see Sec. 3 in \cite{driven}), and we obtain
\begin{eqnarray}
\dot{\rho} &\approx& -\frac{i\omega_{c}}{4}\left[ 2\sigma_1^x\sigma_2^x + \sigma_1^y\sigma_2^y + \sigma_1^z\sigma_2^z, \rho \right] + \nonumber \\
&&\frac{\Gamma}{8} \sum_{j=1,2} \left( \sigma_j^+\rho\sigma_j^+ + \sigma_j^-\rho\sigma_j^- + \sigma_j^z\rho\sigma_j^z - \rho \right) \nonumber \\
&& + \frac{3\Gamma}{16} \sum_{j=1,2} \left( 2\sigma_j^-\rho\sigma_j^+ + 2\sigma_j^+\rho\sigma_j^- \right. \nonumber \\
&& \left.- \sigma_j^+\sigma_j^-\rho - \sigma_j^-\sigma_j^+\rho - \rho\sigma_j^+\sigma_j^- - \rho\sigma_j^-\sigma_j^+ \right) . \label{eq_master_eq_in_ip}
\end{eqnarray}

As already mentioned above, we will study the so-called \lq\lq{\sf X}\rq\rq -states, which have density matrices in the form \cite{Yu2}
\begin{equation}
\rho_S(0) = \rho_{rf}(0) = \rho (0) =
\left( \begin{array}{cccc}
        a_0 & 0 & 0 & w_0 \\
        0 & b_0 & z_0 & 0 \\
        0 & z_0^\ast & c_0 & 0 \\
        w_0^\ast & 0 & 0 & d_0
\end{array} \right) . \label{eq_initial_state}
\end{equation}
A first observation is that this form is not preserved by evolution with Eq. (\ref{eq_master_equation_in_rf}), due to the existence of driving fields; thus the secular approximation brings in a welcome simplification, since it eliminates these fields; indeed, the evolution under Eq. (\ref{eq_master_eq_in_ip}) preserves the density matrix in the \lq\lq{\sf X}\rq\rq form.

We can then take for the structure of the density matrix at any time $t$,
\begin{equation}
\rho(t) =
\left( \begin{array}{cccc}
        a(t) & 0 & 0 & w(t) \\
        0 & b(t) & z(t) & 0 \\
        0 & z^\ast(t) & c(t) & 0 \\
        w^\ast(t) & 0 & 0 & d(t)
\end{array} \right) , \label{eq_x_state}
\end{equation}
and substitute it into (\ref{eq_master_eq_in_ip}) to obtain the following kinetic equation for the non-zero density matrix elements:
\begin{eqnarray}
&& \frac{d}{dt}a(t) = \frac{3}{8}\Gamma [b(t)+c(t)-2a(t)] + \frac{i}{4}\omega_{c}[w(t)-w^\ast(t)], \nonumber \\
&& \frac{d}{dt}b(t) = \frac{3}{8}\Gamma [a(t)-2b(t)+d(t)] + \frac{i}{4}\omega_{c}[z(t)-z^\ast(t)], \nonumber \\
&& \frac{d}{dt}c(t) = \frac{3}{8}\Gamma [a(t)-2c(t)+d(t)] - \frac{i}{4}\omega_{c}[z(t)-z^\ast(t)] , \nonumber \\
&& \frac{d}{dt}d(t) = \frac{3}{8}\Gamma [b(t)+c(t)-2d(t)] - \frac{i}{4}\omega_{c}[w(t)-w^\ast(t)] , \nonumber \\
&& \frac{d}{dt}z(t) = \frac{1}{8}\Gamma [w(t)+w^\ast8t)-10z(t)] + \frac{3i}{4}\omega_{c}[b(t)-c(t)] , \nonumber \\
&& \frac{d}{dt}w(t) = \frac{1}{8}\Gamma [z(t)+z^\ast(t)-10w(t)] + \frac{i}{4}\omega_{c}[a(t)-d(t)] . \nonumber \label{eq_kinetic_eq}
\end{eqnarray}

This system of equations can be analytically solved for arbitrary initial {\sf X} states, and, the concurrence can be computed by a
simplified formula \cite{Yu2}
\begin{equation}
{\cal C}[\rho](t)= 2 \max \left\{ 0, F(t), G(t) \right\} , \label{eq_conc_for_x_state}
\end{equation}
where $F(t) = |z(t)|-\sqrt{a(t)d(t)}$ and $G(t) = |w(t)| -\sqrt{b(t)c(t)}$.
In the following sections, we will study the concurrence evolution for some specific initial {\sf X} states.
To clearly see the limits of  the secular approximation, for all the plots below we have taken  $\Omega=25\Gamma$, which does not strongly satisfy
$\Omega\gg \Gamma$. Otherwise, for larger values of $\Omega$, the secular approximation gives results indistinguishable from numerical simulations.

From now on, we introduce the standard (exponential) decay amplitude $\eta (t) =\exp (-\Gamma t /2)$,  $0\leq \eta \leq 1$, a notation which will be used throughout the remaining sections of the paper.

\section{Werner states}

The Werner state is defined as \cite{brassard}
\begin{eqnarray}
\rho_{W} &=& \frac{1-f}{3}I_{4} + \frac{4f-1}{3}|\Psi^{-}\rangle \langle\Psi^{-}| \nonumber \\
&=& \frac{1}{3}
\left( \begin{array}{cccc}
        1-f & 0 & 0 & 0 \\
        0 & (1+2f)/2 & (1-4f)/2 & 0 \\
        0 & (1-4f)/2 & (1+2f)/2 & 0 \\
        0 & 0 & 0 & 1-f
\end{array} \right) , \label{eq_Werner_state}
\end{eqnarray}
where $|\Psi^{-}\rangle = 1/\sqrt{2}(|01\rangle-|10\rangle)$ is the Bell singlet state.
The parameter $f$ is called fidelity and takes values between $1/4\leq f\leq 1$. Werner
states with fidelity $f\leq 1/2$ are separable;  for $f>1/2$ they have a concurence of $2f-1$.
At zero driving and zero coupling, it is known that there is sudden death only in the region $f\leq 0.714$  and exponential decay for $f >0.714$ \cite{Yu2}. In the case of finite driving, it has been shown that Werner states will all decay by sudden death, as a  consequence of the fact that driving modifies the structure of the decay Liouvillean \cite{driven}. As we will see below, also in the case in which both the coupling and the driving are non-zero, this property remains valid, and all the states decay by sudden death.

\begin{figure}[htb]
\includegraphics[width=7.5cm]{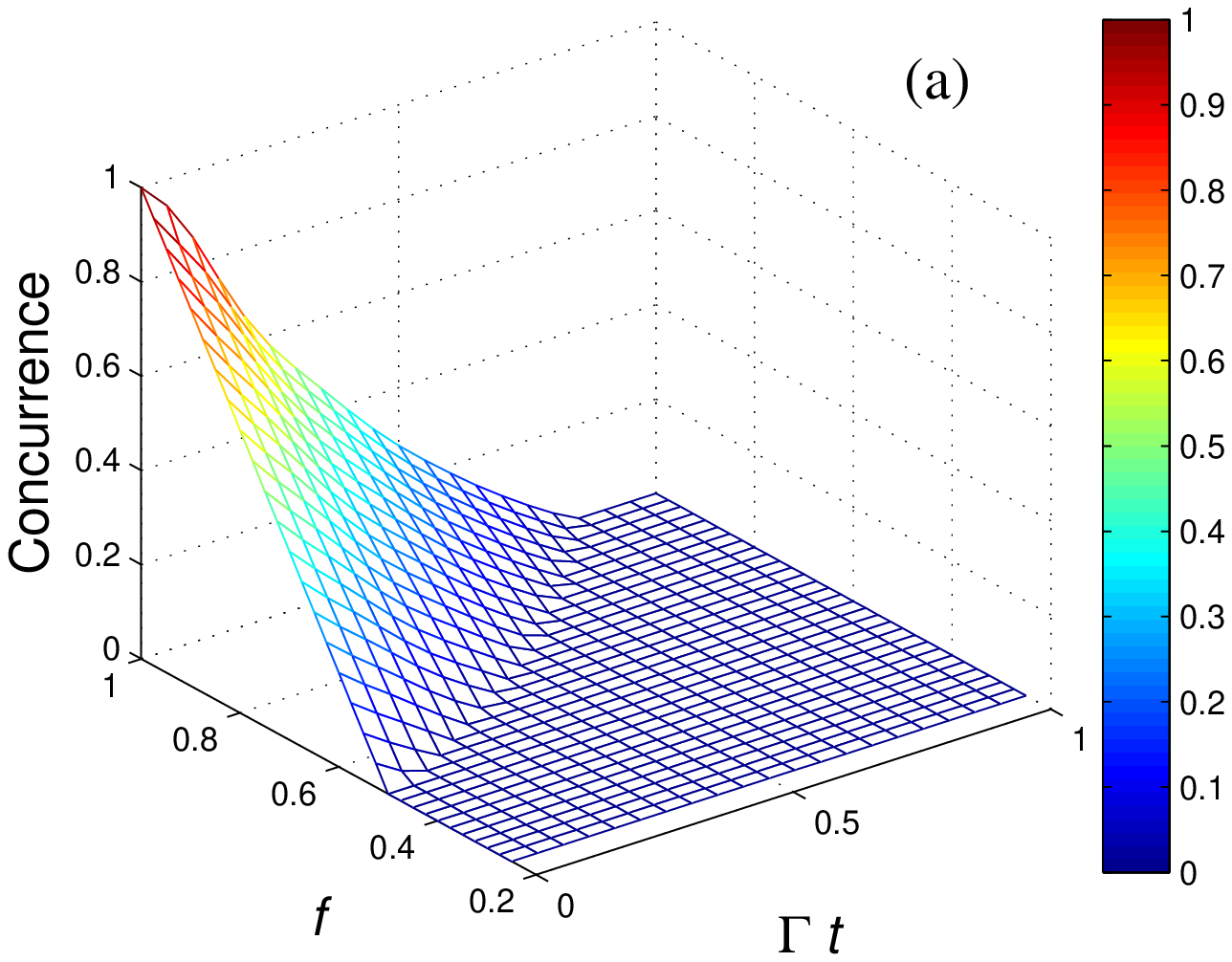}
\includegraphics[width=7cm]{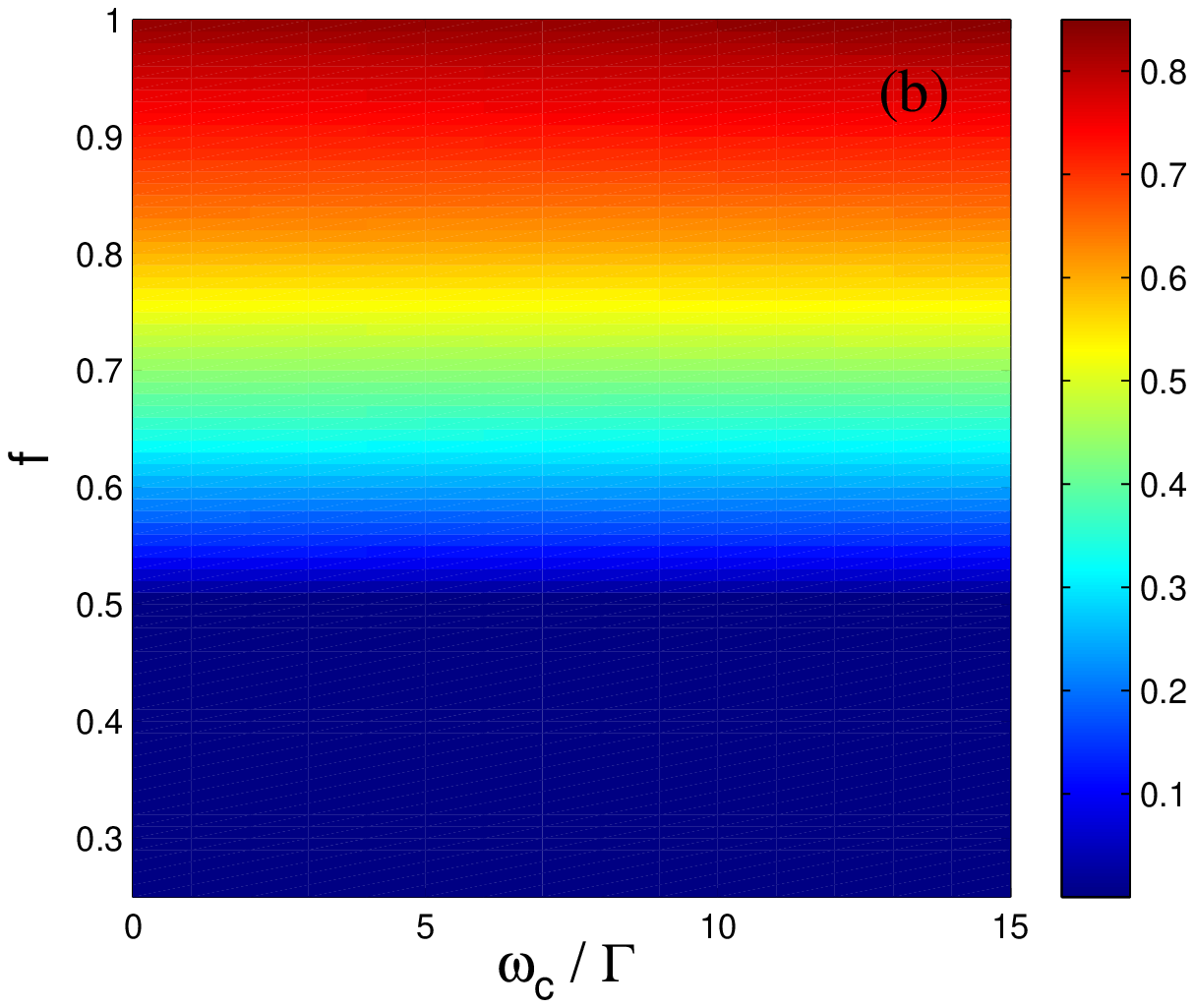}
\caption{(color online) (a) Time evolution of the concurrence for the Werner state, with coupling $\omega_{c} = 5\Gamma$. (b)
Time of sudden death of entanglement as a function of the Werner parameter $f$ of the initial state and the coupling $\omega_{c}$ for $\Omega=25\Gamma$. This plot demonstrates numerically
the analytical result that the time-evolution of Werner states does not depend on coupling. For both simulations, $\Omega = 25\Gamma$. }
\label{werner}
\end{figure}

For the Werner state, Eq. (\ref{eq_kinetic_eq}) has the solution
\begin{eqnarray}
a(t) &=& d(t) = \frac{1}{12}\left[ 3 + (1-4f)\eta^3(t)\right]  , \nonumber \\
b(t) &=& c(t) = \frac{1}{12}\left[ 3 + (4f-1)\eta^3(t)\right] , \nonumber \\
z(t) &=& \frac{4f-1}{12}\eta^2(t)\left[1+ \eta(t) \right], \nonumber \\
w(t) &=& -\frac{4f-1}{12}\eta^3(t)\left[ 1-\eta(t)\right]. \nonumber
\end{eqnarray}
This solution is independent of the qubit-qubit coupling $\omega_{c}$, which means that the enhancement of entanglement
sudden death is also independent of $\omega_{c}$.  This is confirmed numerically
in Fig. \ref{werner}, where we present also a typical time evolution with finite qubit-qubit coupling for these states.
We then obtain the following results for the expression appearing in the formula for the concurrence of X-states, Eq. (\ref{eq_conc_for_x_state}):
\begin{eqnarray}
F(t) &=& \frac{4f-1}{6}\eta^3(t) + \frac{4f-1}{12}\eta^2(t) -\frac{1}{4} , \label{eq_F_Werner} \\
G(t) &=& -\frac{4f-1}{2}\eta^2(t)- \frac{1}{4} . \label{eq_G_Werner}
\end{eqnarray}

Eq. (\ref{eq_G_Werner}) is negative for all values of $f$, therefore the concurrence is determined by Eq. (\ref{eq_F_Werner}): all the states with $f\geq 1/2$ display sudden death due to the fact that the equation $F(t)=0$ has one real root. For example, the maximally entangled state $f=1$ has a sudden death time $t_{s} \approx 0.84 / \Gamma$.

\section{YE state}

The YE states have been introduced in \cite{Yu1},
\begin{equation}
\rho_{YE} = \frac{1}{3}
\left( \begin{array}{cccc}
        1-\alpha & 0 & 0 & 0 \\
        0 & 1 & 1 & 0 \\
        0 & 1 & 1 & 0 \\
        0 & 0 & 0 & \alpha
\end{array} \right) , \label{eq_YE_state}
\end{equation}
where the parameter $\alpha$ takes values in the interval $[0,1]$.

In the case of zero coupling and  zero driving it has been shown \cite{Yu1} that  only the states with $\alpha > 1/3$ suffer sudden death on entanglement, while the states with $\alpha <1/3$ decay exponentially. For finite driving and zero coupling all the YE states
decay by sudden death \cite{driven}, and this feature is preserved also in the case of both nonzero coupling and driving, as we will see below.

With this initial state, the solution of Eq. (\ref{eq_kinetic_eq}) is
\begin{eqnarray}
a(t) &=& \frac{1}{12} \left\{ 3 - \eta^3(t) - 2(2\alpha-1)\left[ \cosh\left(\frac{\zeta t}{4}\right) \right.\right.\nonumber \\
&& \left.\left.+ \sinh\left(\frac{\zeta t}{4}\right)\frac{\Gamma}{\zeta} \right] \eta^2(t) \right\} , \nonumber \\
d(t) &=& \frac{1}{12} \left\{ 3 - \eta^3(t) + 2(2\alpha-1)\left[ \cosh\left(\frac{\zeta t}{4}\right)  \right.\right.\nonumber \\
&&\left.\left.+\sinh\left(\frac{\zeta t}{4}\right)\frac{\Gamma}{\zeta} \right] e^{-\Gamma t} \right\} ,\nonumber \\
b(t) &=& c(t) = \frac{1}{12}\left[ 3 + \eta^3(t)\right] , \ \ \ \ z(t)=  \frac{1}{6}\eta^2(t)\left[ 1+ \eta(t)\right] , \nonumber \\
 w(t) &=& \frac{1}{3}\left[ -i(2\alpha -1)\frac{\omega_{c}}{\zeta}\sinh\left( \frac{\zeta t}{4}\right) +\frac{1}{2}\left(1- \eta(t)\right) \right]  \eta^2(t) ,\nonumber
\end{eqnarray}
with
\begin{equation}
\zeta \equiv \sqrt{\Gamma^2 - 4(\omega_{c})^2}. \nonumber\label{eq_zeta}
\end{equation}

In the limit of $\omega_{c}\gg \Gamma$, we can approximate $\zeta \approx 2 i \omega_{c}$, and we find
\begin{eqnarray}
F(t) &\approx& - \frac{1}{12}\sqrt{[3-\eta^3(t)]^2 - 4 (2\alpha - 1)^2 \eta^2(t)\cos^2\left(\frac{\omega_{c}t}{2}\right)} ´\nonumber \\
&& +\frac{\eta^3(t)+\eta^2(t)}{6} , \label{eq_F_YE} \\
G(t) &\approx& \frac{\eta^2(t)}{6}\sqrt{(2\alpha - 1)^2 \sin^2\left(\frac{\omega_{c}t}{2}\right) + (1-\eta(t))^2} \nonumber \\
&& - \frac{3+\eta^3(t)}{12} . \label{eq_G_YE}
\end{eqnarray}

\begin{figure}[htb]
\includegraphics[width=8cm]{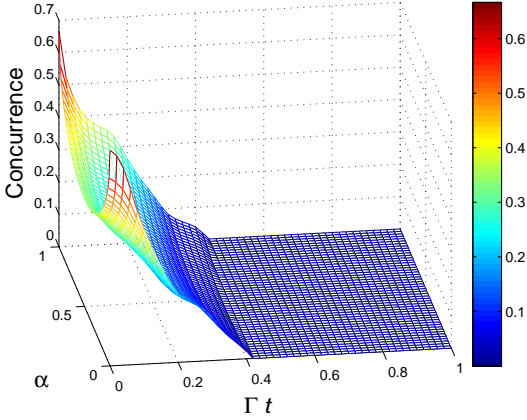}
\caption{(color online) Time evolution of the concurrence for the YE state, with coupling $\omega_{c} = 5\Gamma$ and $\Omega=25\Gamma$. }
\label{timedynamics_YE}
\end{figure}

\begin{figure}[htb]
\includegraphics[width=7cm]{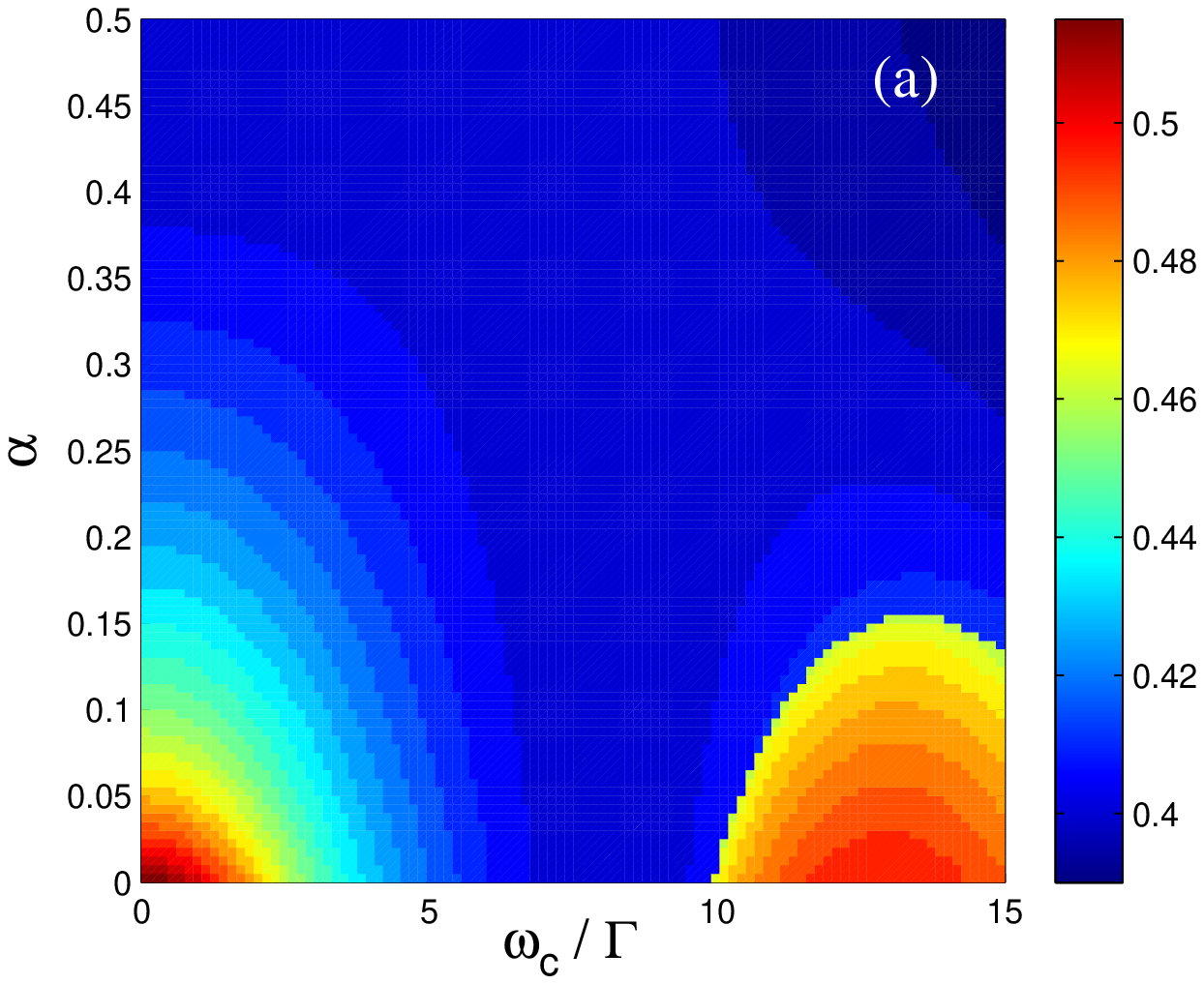}
\includegraphics[width=7cm]{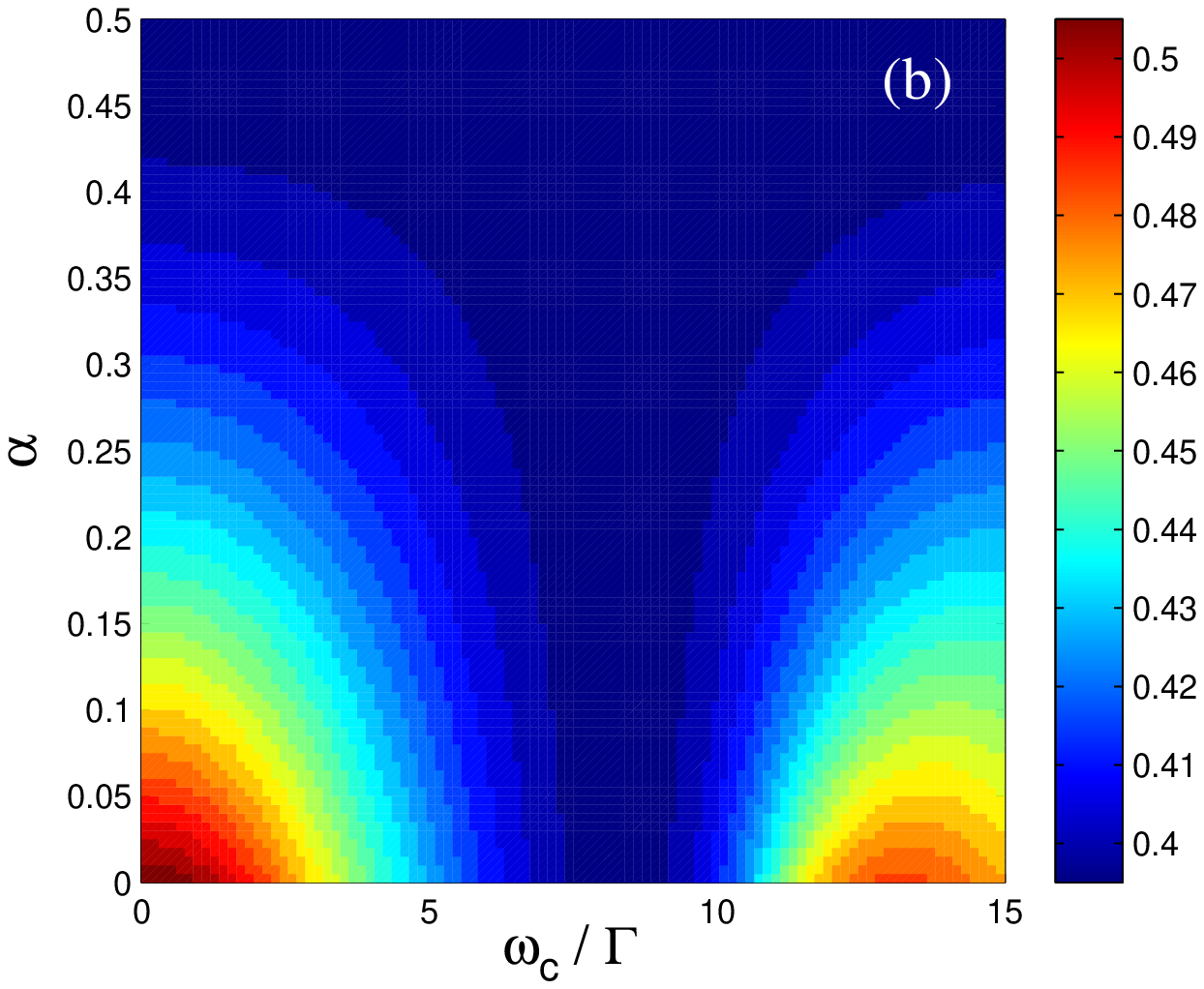}

\caption{(color online)  (a) Time of sudden death of entanglement of YE states as a function of the Werner parameter $\alpha$ of the initial state and the coupling $\omega_{c}$ for $\Omega=25\Gamma$, obtained numerically. (b) The corresponding analytical result of (a) in the secular approximation.
}
\label{diagram_YE}
\end{figure}

We note that $G(t)$ is negative for all values of $\alpha$, and therefore the equation $F(t)=0$ sets the value
of $t_{ESD}$.
A typical example of evolution is shown in Fig. \ref{timedynamics_YE}. These states start with nonzero entanglement, which decays to zero in an time $t_{ESD}$ shown in the parameter space $s - \omega_{c}$ in Fig. \ref{diagram_YE}. However, unlike the diagonal states that we will analyze below, the YE state displays  only very mild delayed birth of of entanglement phenomena at very large $\omega_{c}/\Gamma$. For moderate values of $\omega_{c}/\Gamma$, the transitions resulting in the lobs of Fig. \ref{diagram_YE} are smooth.

\section{Diagonal states}

An important class of states are the diagonal states, of which we consider two subclasses, namely states that are diagonal on a subspace of single excitations ({\it i.e.} only one quanta is present initially in the system), and states which are diagonal in the subspace with zero and two excitations. Since $g$ and $e$ are typically used as
indices for ground, respectively excited states, we will call $\rho_{eg-ge}$ the first class of states and $\rho_{ee-gg}$ the second class. Both these states have zero initial entanglement: as we will see however, a finite concurrence is acquired during the short-term evolution and decays in a finite time afterwards.

\subsection{Diagonal one-excitation states ($eg-ge$ states)}
\label{eggesection}

The generic form of $eg-ge$ states is
\begin{equation}
\rho_{eg-ge} = (1-p)|10\rangle\langle 10\rangle + p |01\rangle\langle 01| =
\left( \begin{array}{cccc}
        0 & 0 & 0 & 0 \\
        0 & 1-p & 0 & 0 \\
        0 & 0 & p & 0 \\
        0 & 0 & 0 & 0
\end{array} \right), \label{eq_1e_state}
\end{equation}
where the parameter $p\in [0, 1]$. The physical picture that these states correspond to  is the following: suppose we have two identical qubits, each with its own source of dissipation. We excite one of them  by a fast $\pi$ pulse (fast compared to $\Gamma$) and then we allow them to interact with a coupling strength $\omega_{c}$. For example, with quantum superconducting circuits this type of initial state is realized in most experiments \cite{direct,indirect}.

\begin{figure}[htb]
\includegraphics[width=8cm]{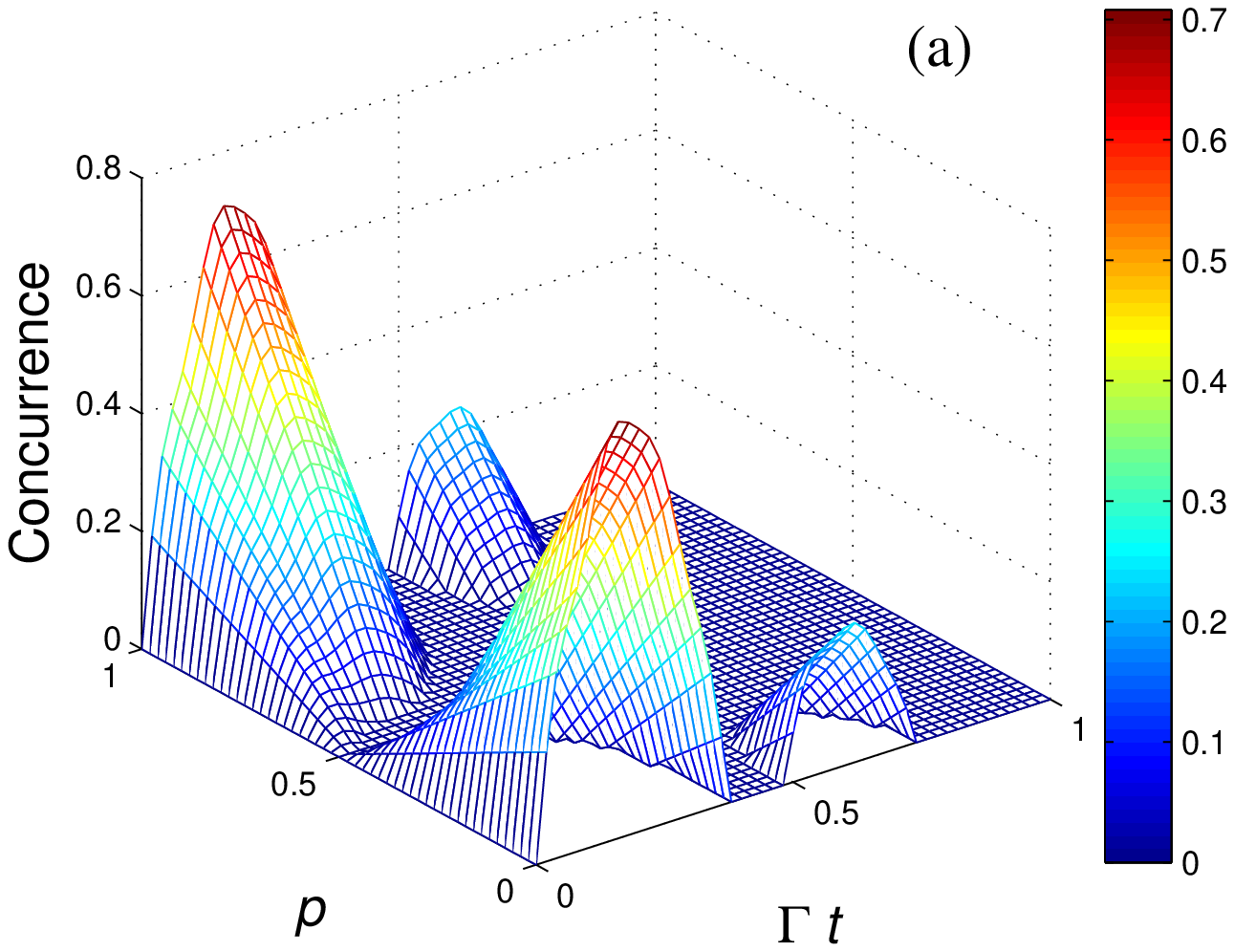}
\includegraphics[width=8cm]{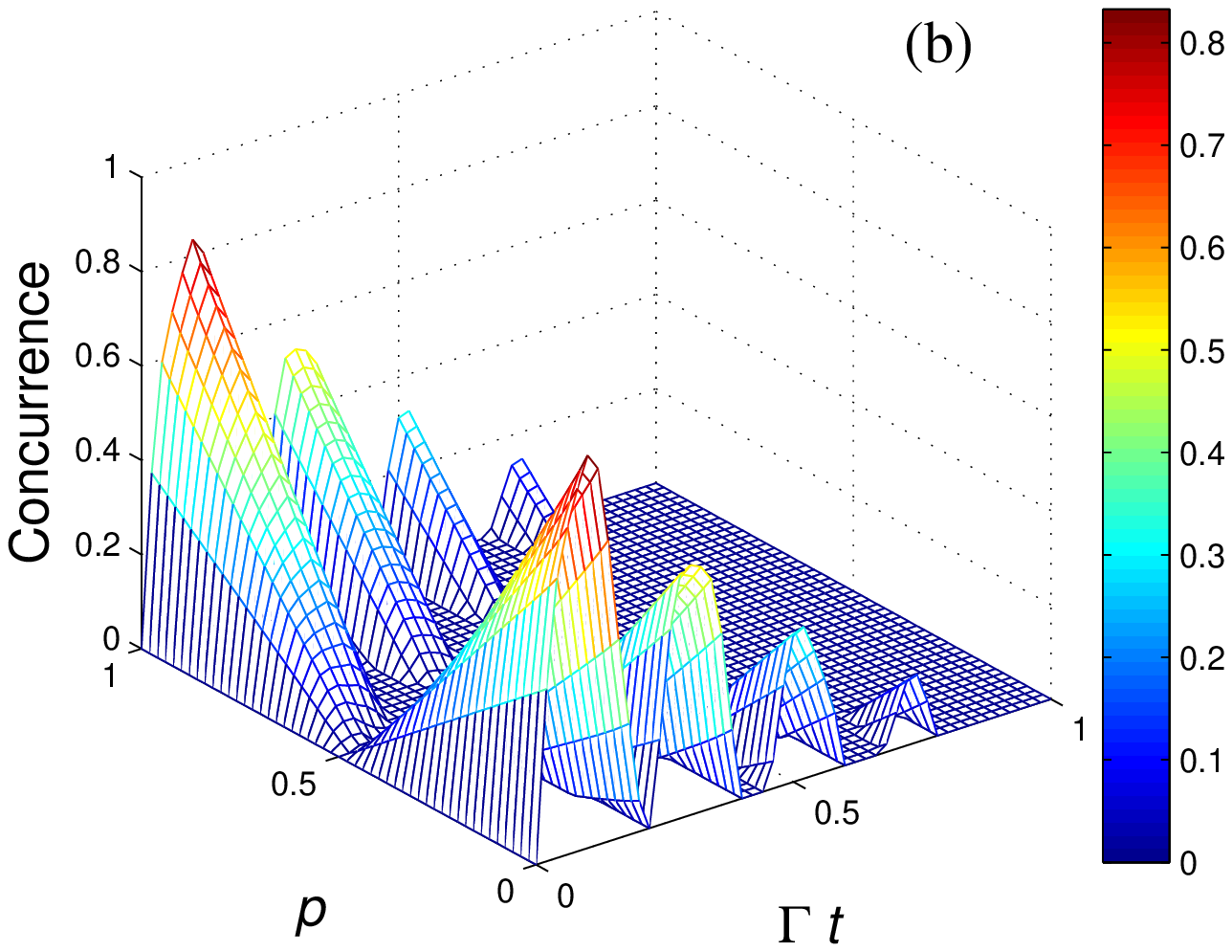}
\caption{(color online) (a) Time evolution of the concurrence for the $eg-ge$ state, with coupling $\omega_{c} = 5\Gamma$ and $\Omega=25\Gamma$. (b)
Time evolution for $\omega_{c} = 10\Gamma$ and $\Omega=25\Gamma$.
}
\label{timedynamics_egge}
\end{figure}

In the absence of  dissipation we would expect the transferring back and forth of a single quanta of excitation between the two qubits, with periodic creation of  entanglement due to coupling. The dynamics of the system would explore only the subspace spanned by $|01\rangle$ and $|10\rangle$, having at times  components on the maximally entangled Bell singlet states
$|\Psi^{\pm}\rangle = (|01\rangle \pm |10\rangle )/\sqrt{2}$. As we will see below, in the presence of dissipation and driving, this periodic process is still present, and if it is faster than the decay rate it leads to periodic creation and extinction of entanglement, as the concurrence competes with the effect of the environmental noise.  Numerical simulations are presented  in  Fig. \ref{timedynamics_egge}.

For $eg-ge$ states, the solution of Eq. (\ref{eq_kinetic_eq}) is
\begin{eqnarray}
a(t) &=& d(t) = \frac{1}{4}\left[ 1 - \eta^3(t)\right]  , \ \ \ \ w(t) = 0,  \nonumber \\
 b(t) &=& \frac{1}{4}\left\{ 1 + \eta^3(t)+ 2 (2 p - 1) \left[ \cosh\left( \frac{\xi t}{4} \right) \right.\right.\nonumber \\
 && \left.\left.+ \sinh\left( \frac{\xi t}{4} \right)\frac{\Gamma}{\xi} \right]\eta^2(t)\right\}  , \nonumber \\
c(t) &=&  \frac{1}{4}\left\{ 1 + \eta^3(t) - 2 (2 p - 1) \left[ \cosh\left( \frac{\xi t}{4} \right) \right.\right. \nonumber \\
&&\left.\left.+ \sinh\left( \frac{\xi t}{4} \right)\frac{\Gamma}{\xi} \right]\eta^2(t)\right\} , \nonumber \\
z(t) &=& \frac{3 i \omega_{c}}{\xi}(2 p - 1)\sinh\left( \frac{\xi t}{4} \right)\eta^2(t) ,\nonumber
\label{egge_analytical}
\end{eqnarray}
with $\xi \equiv \sqrt{\Gamma^2 - 36(\omega_{c})^2}\approx 6i\omega_{c}$, for $\omega_{c}\gg \Gamma$.

\begin{figure}[htb]
\includegraphics[width=7cm]{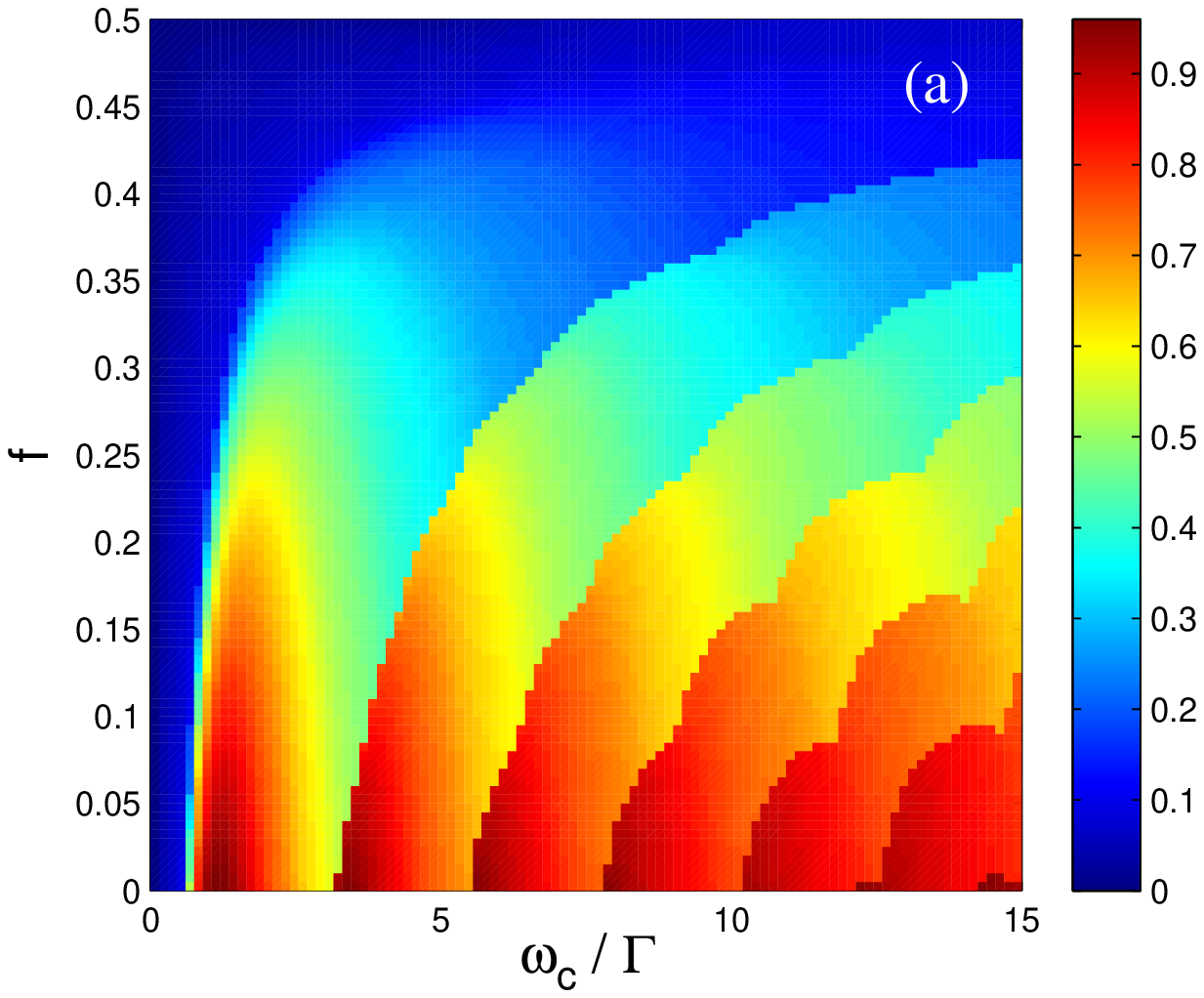}
\includegraphics[width=7cm]{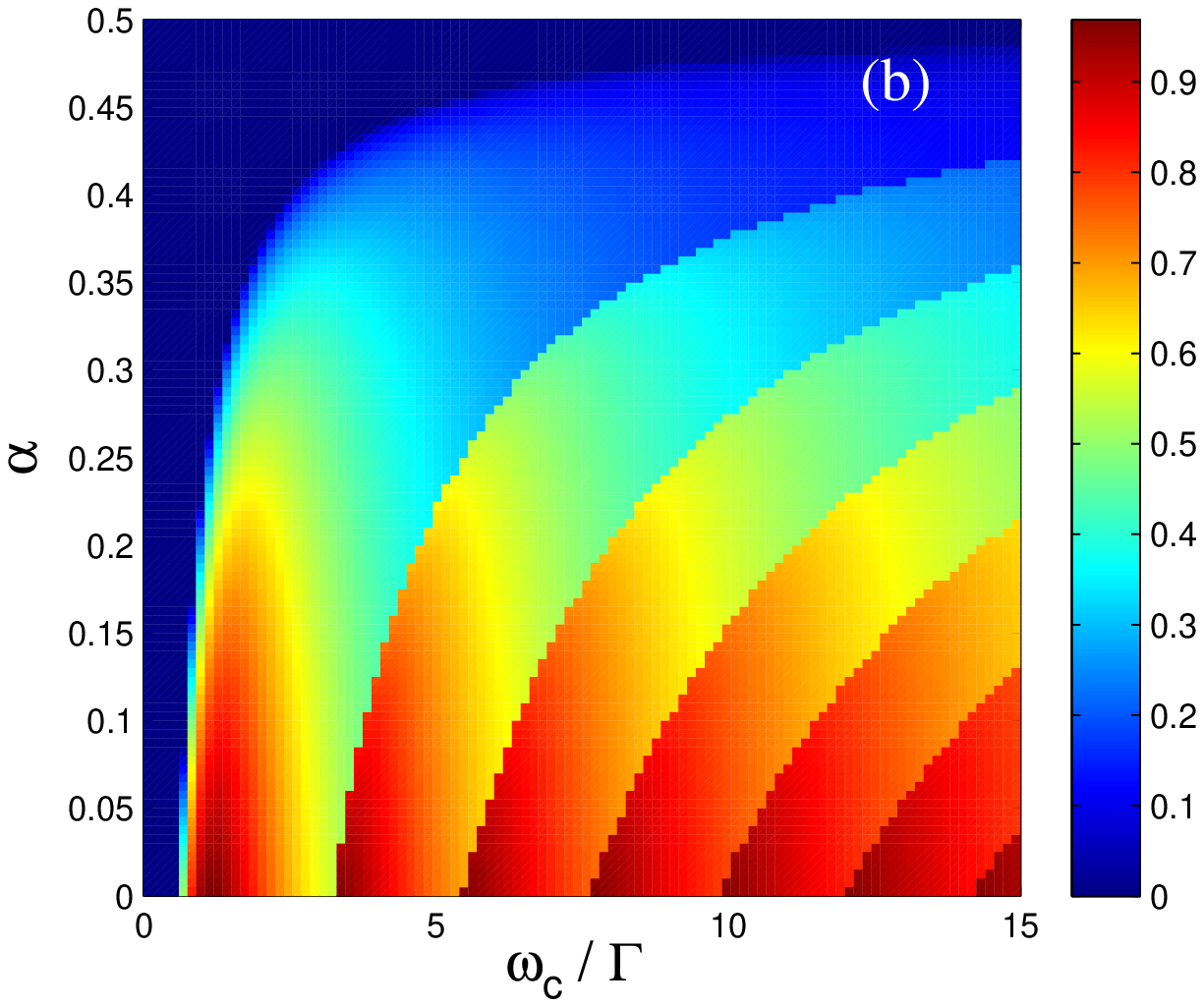}

\caption{(color online)  (a) Time of sudden death of entanglement for $eg-ge$ states as a function of parameter $a$ of the initial state and the coupling $\omega_{c}$ for $\Omega=25\Gamma$, obtained numerically. (b) The corresponding analytical result in the secular approximation.
}
\label{diagram_egge}
\end{figure}

Since now $w = 0$, the concurrence simplifies to
\begin{equation}
{\cal C}(\rho) = 2 \max \left\{ 0, F(t) \right\} ,
\end{equation}
where
\begin{equation}
F(t) \approx \frac{\eta^2}{2}\left| (2p - 1) \sin\left( \frac{3\omega_{c}t}{2} \right) \right| - \frac{1}{4} \left[ 1-\eta^3(t) \right] , \label{eq_F_1E}
\end{equation}
which is shown in  Fig. \ref{diagram_egge}(b). We notice here that the initial state has zero entanglement: later, some entanglement builds up, decays, and  revives after some time. From Eq. (\ref{eq_F_1E}) we see that there is indeed a periodic process (showing up in the sine function) which, if it is fast enough,  drives  the function $F$ to positive values. The period of this process is independent on $p$ and has the value $2 \pi /3\omega_{c}$, which agrees very well with the period of the maxima extracted from numerical simulations such as Fig. \ref{timedynamics_egge}. Of course, in the absence of dissipation and driving, the concurrence reduces to the
expected purely oscillatory value ${\cal C} = (2p-1)|\sin (3 \omega_{c}t /2 )|$.

Fig. \ref{diagram_egge} shows  some unusual features, seen as sudden changes of color along the axis $\omega_{c}$, which mark the appearance of another lob structure.
When calculating $t_{ESD}$, the code evolves the system in time, and records the last time value for which the concurrence was positive. We have verified numerically that the sudden shifts in $t_{ESD}$ correspond to the appearance  of a new entanglement revival (of the type shown in Fig. \ref{timedynamics_egge}). The evolution within the $k$th lobe in diagram Fig. \ref{diagram_egge} corresponds to $k$ delayed entanglement sudden birth events (for example,  the value  $\omega_{c} = 10\Gamma$ and $\Omega=25\Gamma$ as in Fig. \ref{timedynamics_egge}(b) corresponds  to the fourth lobe in Fig. \ref{diagram_egge}).

\subsection{Diagonal $ee-gg$ states}

We now consider states parametrized by $s\in [0,1]$ of the type
\begin{equation}
\rho_{ee-gg} = s|11\rangle\langle 11\rangle + (1-s)|00\rangle \langle 00| =
\left( \begin{array}{cccc}
        s & 0 & 0 & 0 \\
        0 & 0 & 0 & 0 \\
        0 & 0 & 0 & 0 \\
        0 & 0 & 0 & 1-s
\end{array} \right).
\end{equation}

Physically, they generalize the situation in which two qubits, both in their excited state, are put in contact.

\begin{figure}[htb]
\includegraphics[width=8cm]{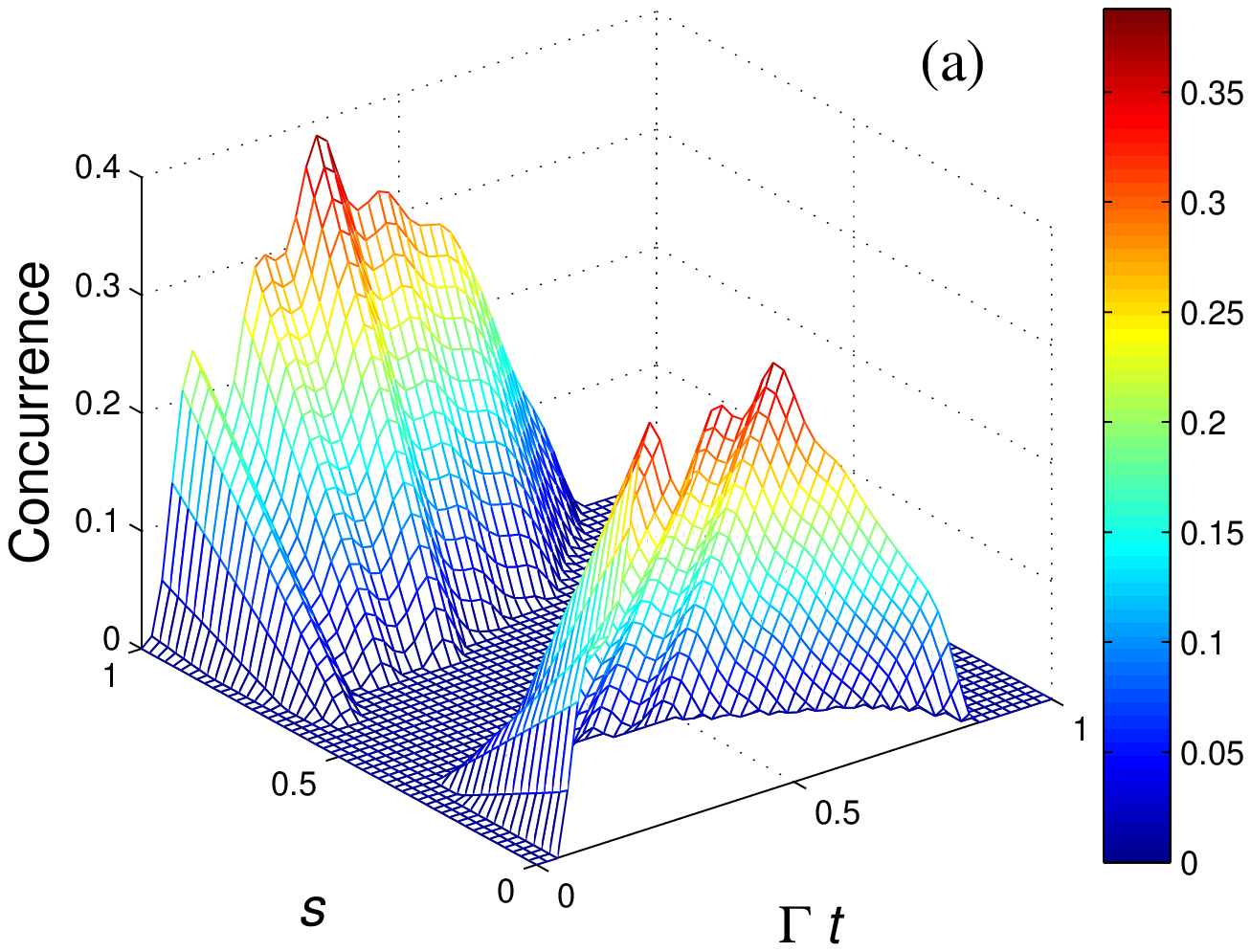}
\includegraphics[width=8cm]{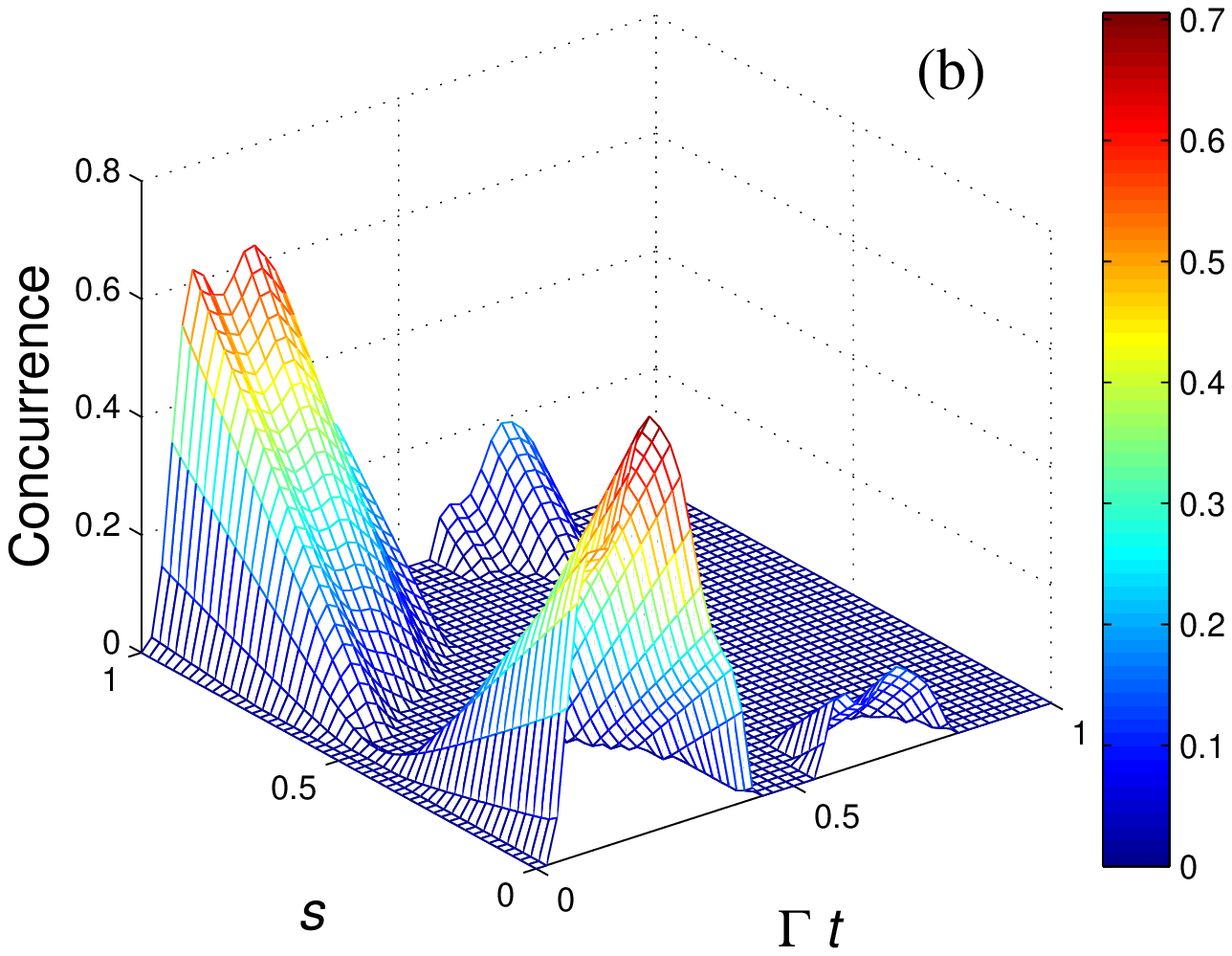}
\caption{(color online) (a) Time evolution of the concurrence for the $ee-gg$ state, with coupling $\omega_{c} = 5\Gamma$ and $\Omega =25\Gamma$.
(b) Time evolution of the concurrence for the $ee-gg$ state, with coupling $\omega_{c} = 13\Gamma$ and $\Omega =25\Gamma$.}

\label{timedynamics_eegg}
\end{figure}

\begin{figure}[htb]
\includegraphics[width=7cm]{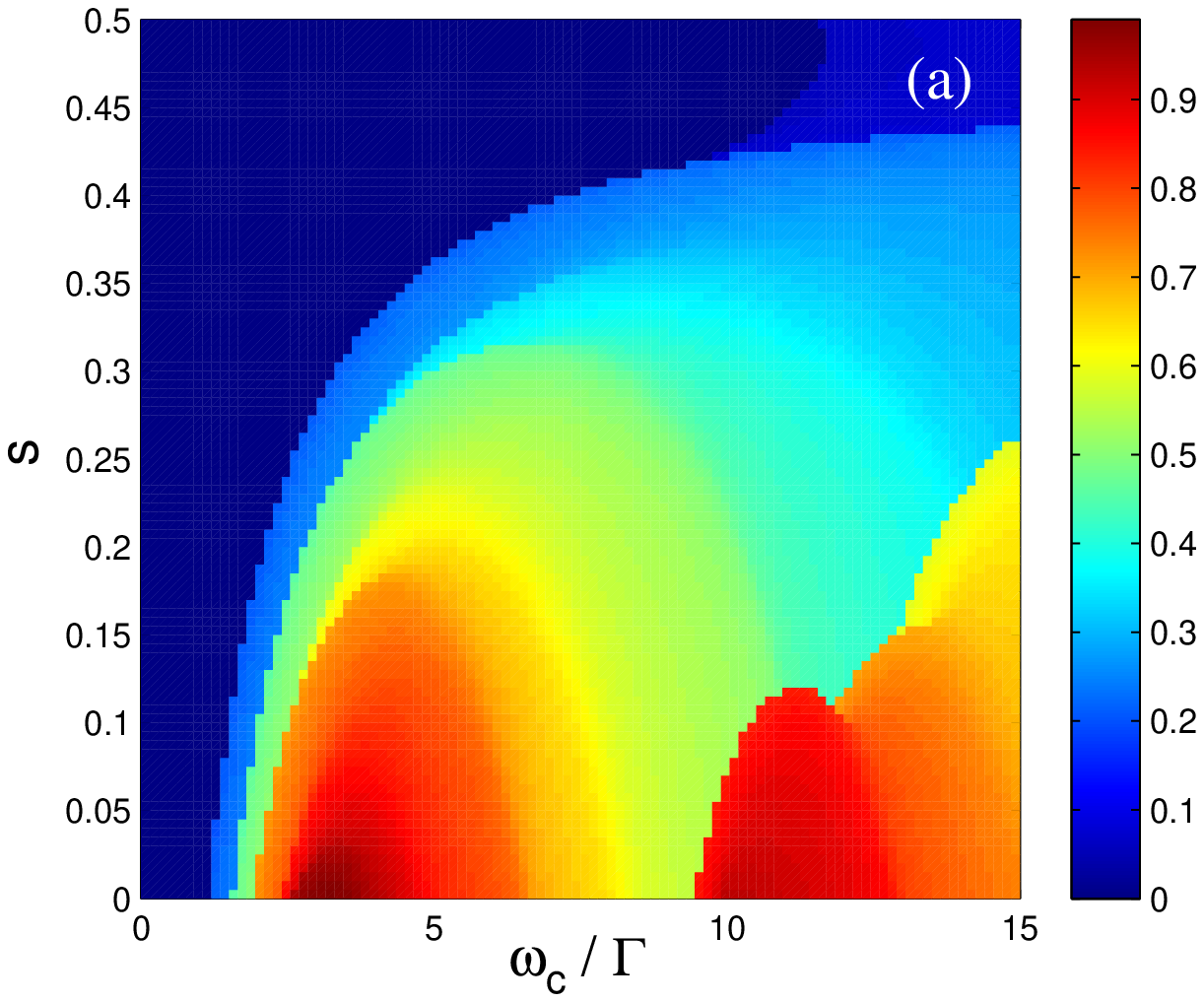}
\includegraphics[width=7cm]{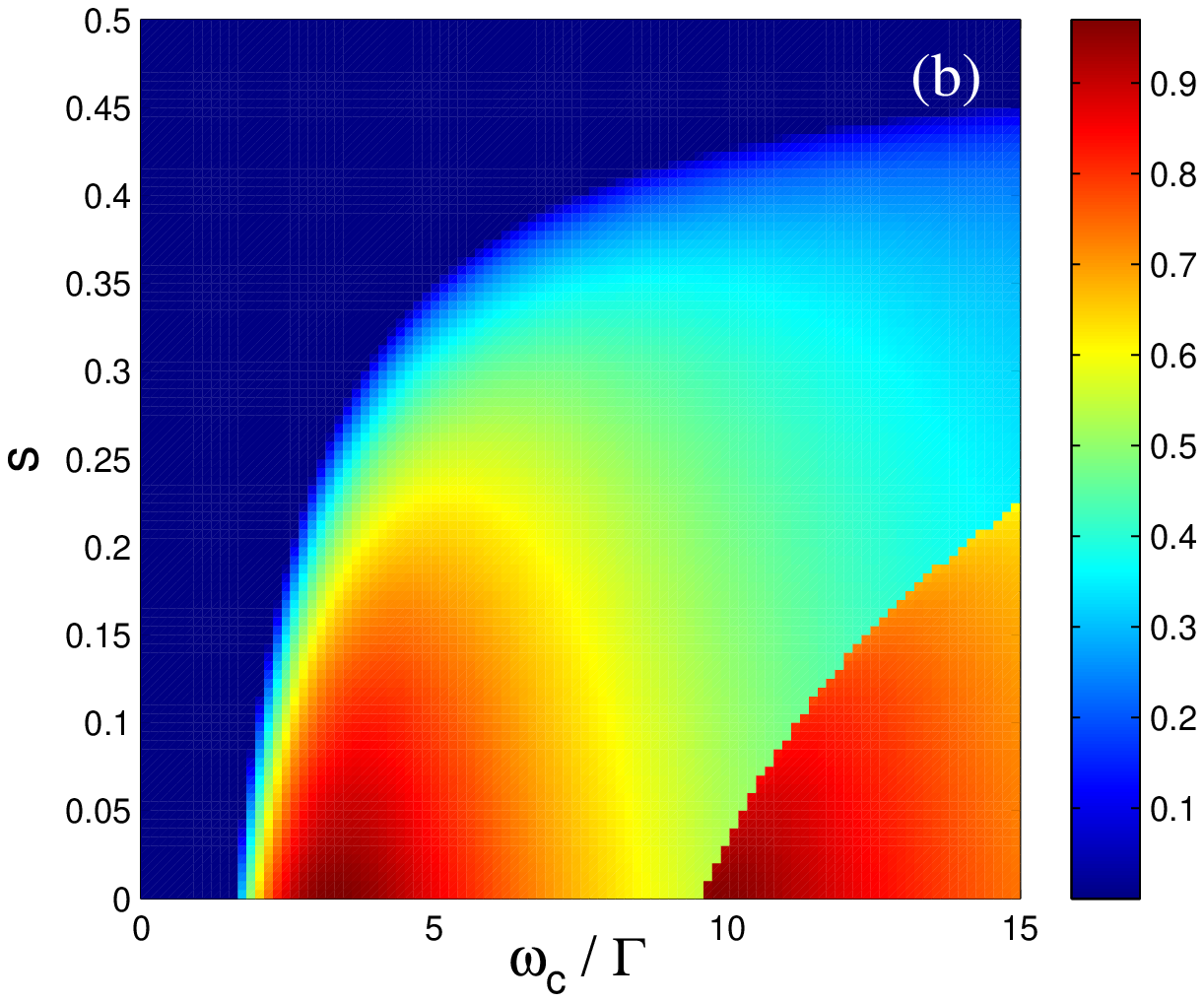}

\caption{(color online)  (a) Time of sudden death of entanglement for $ee-gg$ states as a function of parameter $s$ of the initial state and the coupling $\omega_{c}$ for $\Omega=25\Gamma$, obtained numerically. (b) The corresponding analytical result of (a) in the secular approximation.
}
\label{diagram_eegg}
\end{figure}

Again, we can solve analytically Eq. (\ref{eq_kinetic_eq}),
\begin{eqnarray}
b(t) &=& c(t)= \frac{1}{4}\left[ 1 - \eta^3(t)\right] , \nonumber \\
w(t) &=& \frac{i(1-2s)\omega_{c}}{\zeta}\sinh\left(\frac{\zeta t}{4} \right)e^{-\Gamma t}, z(t) = 0 ,\nonumber \\
a(t) &=& \frac{1}{4}\left\{ 1 +  \eta^3(t)+ 2 (2s - 1) \left[ \cosh\left( \frac{\xi t}{4} \right) \right.\right. \nonumber \\
&& \left.\left.+ \sinh\left( \frac{\xi t}{4} \right)\frac{\Gamma}{\xi} \right]\eta^2(t) \right\}  , \nonumber \\
d(t) &=&  \frac{1}{4}\left\{ 1 +  \eta^3(t) - 2 (2s - 1) \left[ \cosh\left( \frac{\xi t}{4} \right) \right.\right. \nonumber \\
&& \left.\left. + \sinh\left( \frac{\xi t}{4} \right)\frac{\Gamma}{\xi} \right]\eta^2(t)\right\} , \nonumber
\label{eq_solution_of_kinetic_eq}
\end{eqnarray}
where $\zeta$ is defined in Eq. (\ref{eq_zeta}). This time, the concurrence (\ref{eq_conc_for_x_state}) is reduced for $z(t)=0$ to
\begin{equation}
{\cal C}(\rho) = 2 \max \left\{ 0, G(t)\right\} . \label{eq_conc_for_uu_state}
\end{equation}
In the limit of $\omega_{c}\gg \Gamma$, we find
\begin{equation}
G(t)\approx \frac{1}{2}\left|(1-2s)\sin \frac{\omega_{c}t}{2}\right| \eta^2(t) - \frac{1}{4}\left[ 1 -  \eta^3(t)\right],
\end{equation}
which describes again an oscillatory process overlapping with decay.
As in the analysis before, $t_{ESD}$ will display  revivals of entanglement with period $2\pi /\omega_{c}$, a statement that we have also checked numerically. In Fig. \ref{timedynamics_eegg} we show
the corresponding diagram of $t_{ESD}$ in the $s-\omega_{c}$ parameter space. As in the case of $eg-ge$ states, the appearance of lobes in this diagram
(sudden large $t_{ESD}$) is due to the emergence of a new entanglement revival. For example, at $\omega_{c} = 13\Gamma$ and $\Omega =25\Gamma$ as in Fig. \ref{timedynamics_eegg}(b), we have two revivals and, with these parameters, we are at the second lobe in Fig. \ref{diagram_eegg}.

\section{Discussion: coupled undrived systems and finite temperature effects}

At this point it is worth asking the following question: is it possible to separate the roles played by the environment, by the driving, and by the coupling? We already know what happens when coupling is not present, as most of the literature on
the topic of sudden death of entanglement studies this situation. But then, if we have coupling, isn't it the case that this will produce anyway some form of entanglement -- then, does the driving play in fact any role in the emergence of the structures for $t_{ESD}$ presented above? The answer is yes. As pointed out already throughout this paper, the existence of the type of effects presented in this paper depends on all three ingredients: driving, coupling, and  bipartite dissipation. Note that bipartite reservoirs are indeed needed: in the case of a single reservoir, the decay is exponential and $t_{ESD}$ is infinite.

Indeed, if one takes the master equation Eq. (\ref{eq_master_equation_in_rf}) and simulates the evolution with $\Omega=0$, the corresponding diagrams have no interesting features: there is, somewhat surprisingly, no dependence of $t_{ESD}$ on $\omega_{xx}$, so none of the diagrams would be obtained (with the exception of that for Werner states, which has the special property of $\omega_{xx}$- independence even at finite driving).

In the following we summarize the results obtained by numerical simulations for the no-driving situation. To make the discussion more general, we will introduce also temperature in the problem,

\begin{eqnarray}
{\cal L}^{(th)}_{j}[\rho] &=& \sum_{j=1,2}\frac{\gamma_j}{2} (\bar{n}_j + 1) \left( 2\sigma_j^-\rho\sigma_j^+ - \sigma_j^+\sigma_j^-\rho - \rho\sigma_j^+\sigma_j^- 
\right) \nonumber \\
&& + \frac{\gamma_j}{2}\bar{n}_j \left( 2\sigma_j^+\rho\sigma_j^- - \sigma_j^-\sigma_j^+\rho - \rho\sigma_j^-\sigma_j^+ \right) ,
\end{eqnarray}
where $\bar{n}_j$ are the bosonic thermal averages,
\begin{equation}
\bar{n}_{j}=[\exp (\hbar\omega_{j}^{L}/k_{B}T )-1]^{-1},
\end{equation}
and we simulate the evolution of the system under
\begin{eqnarray}
\dot{\rho}_S = -i [H_{sys}, \rho_S] +  {\cal L}^{(th)}[\rho_S] .
\end{eqnarray}
with $\Omega_1 =\Omega_2 = 0$.

We start first with the observation that, in the general case in which we have driving, coupling, and decoherence in two reservoirs at finite temperature, the only effect of the temperature is to blur the diagrams obtained before, as intuitively expected.
Also, for $\Omega = 0$ and at zero temperature, the problem can be solved analytically \cite{agarwal}. Here we will show that finite temperature can have nontrivial effects, as demonstrated by the case of  $eg-ge$ states.

For Werner states, as expected, there is no dependence on $\omega_{c}$, either at zero temperature or at finite temperature.
Also here, the only effect of temperature is that it decreases the sudden death time and tends to transform also the
regions which were decaying exponentially (e.g. for Werner states, for $0.714<f<1$ \cite{Yu2}) into sudden death.

For YE states, we find numerically that there is no change of concurrence as a function of $\omega_{c}$, either at zero temperature or at finite temperature. The only effect of finite temperature is that it makes the YE states that would otherwise decay exponentially at zero temperature (the states with $1/4<\alpha <1/3$) become sudden-death states as well.

For $ee-gg$ states $t_{ESD}$ is zero, as the initial state has no entanglement: the coupling does not create
entanglement by itself. Pumping is needed here. There is no interesting effect at finite temperature: $t_{ESD}$ remains zero.

\begin{figure}[htb]
\includegraphics[width=8cm]{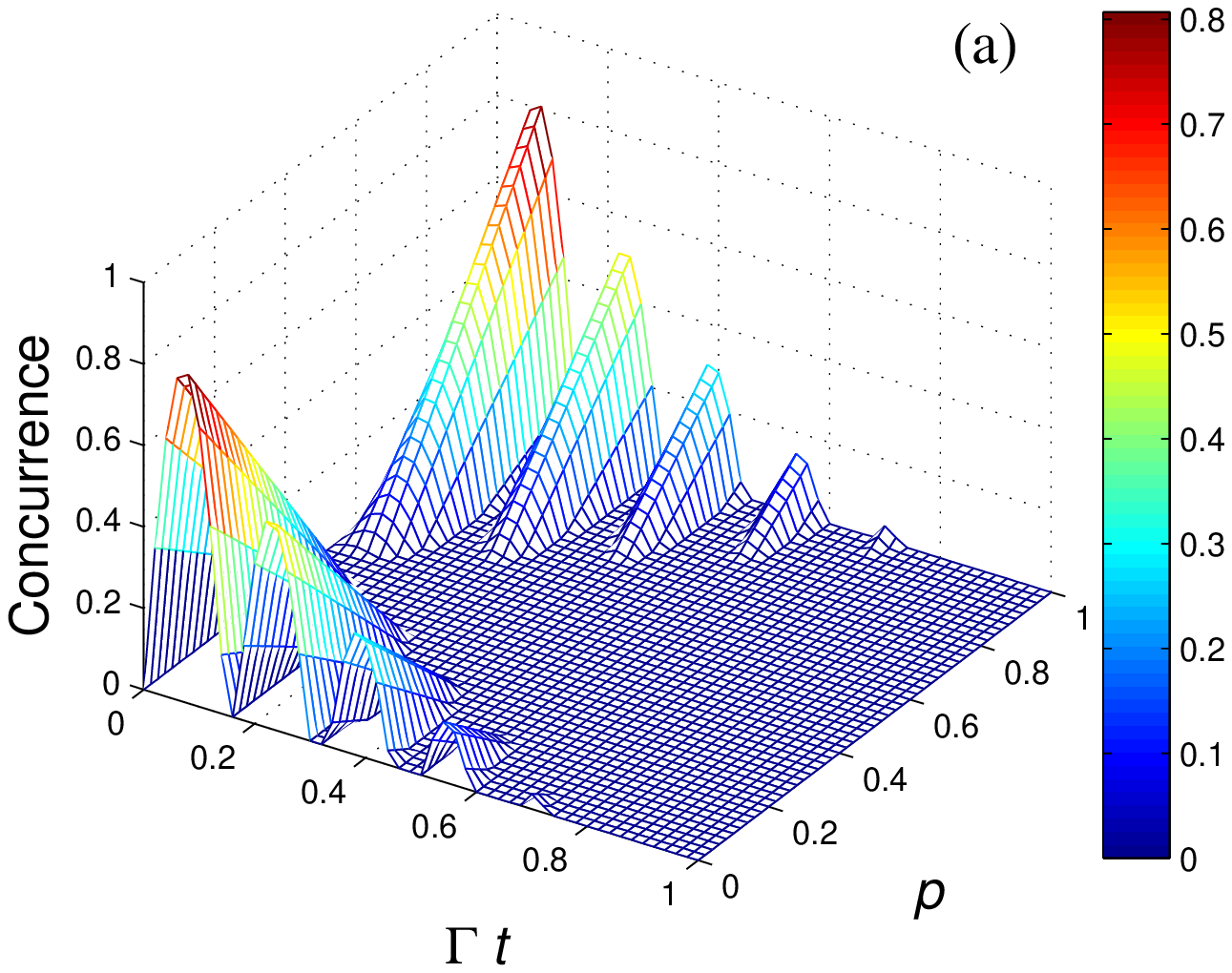}
\includegraphics[width=7.5cm]{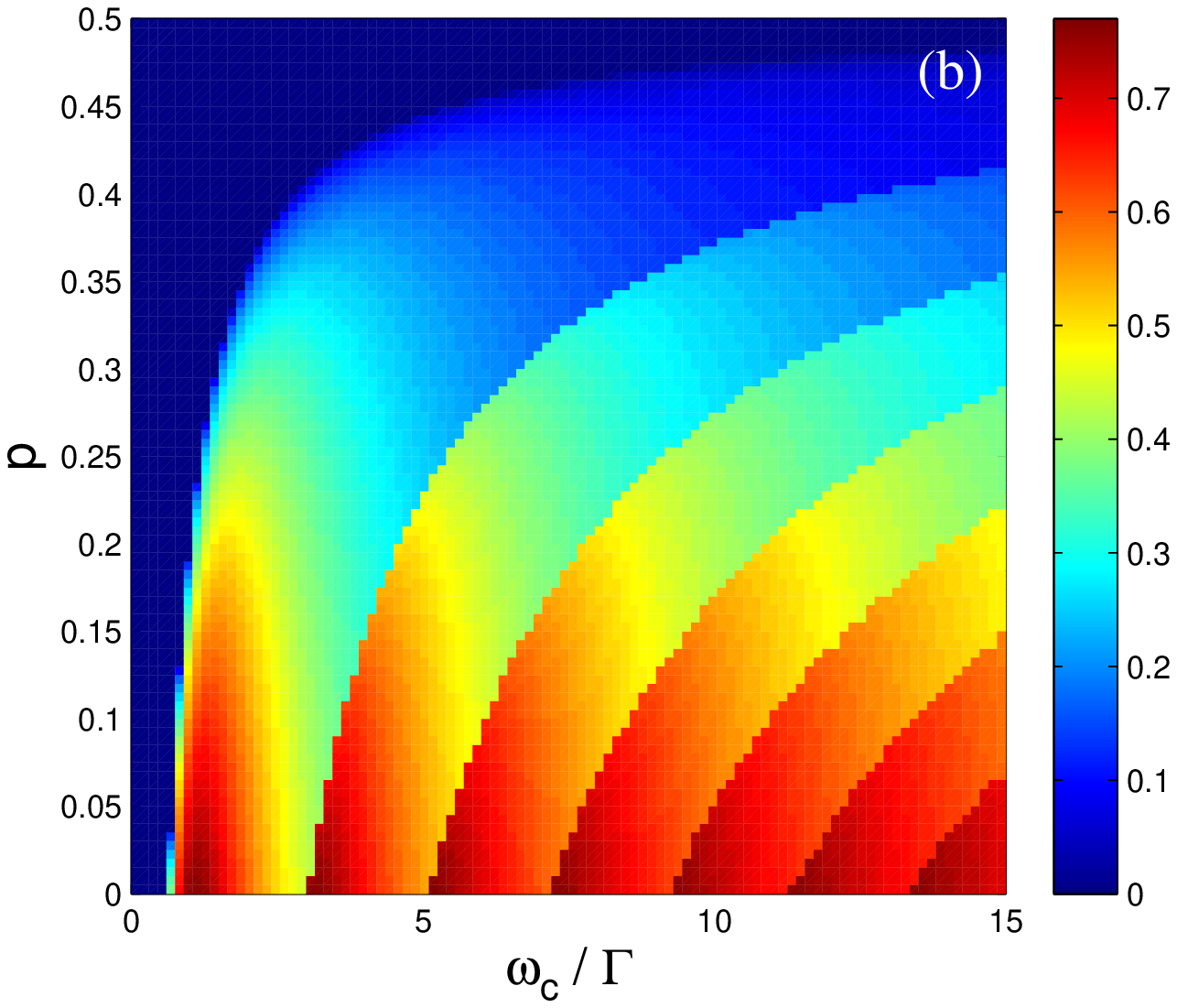}
\caption{(color online) (a) Time evolution of the concurrence for  $eg-ge$ states, with coupling $\omega_{c} = 10\Gamma$ and $\Omega_{1} = \Omega_{2} = 0$.
(b) $t_{ESD}$ for $eg-ge$ states. For both plots the temperature corresponds to average bosonic occupation numbers $\bar{n}_{1}=\bar{n}_{2}=0.25$.}
\label{egge_temperature}
\end{figure}

For  $eg-ge$  states at zero temperature there is no structure in the $t_{ESD}$ diagram, again due to the fact that the
states decay exponentially. Also at very large temperatures ($n\gg1 $) there is no structure: this time because the decay is too fast and $t_{ESD}$ becomes essentially zero. However, for intermediate temperatures a structure somewhat similar with that obtained in the case of driving emerges, shown in  Fig. \ref{egge_temperature} (b) for $\bar{n}_{1}=\bar{n}_{2}=0.25$.

This is somehow unexpected: usually the situation is that features appearing at a certain temperature would be enhanced at even lower temperatures (or in other words the only effect of temperature is usually to wash out the features seen at zero or very low temperatures).  We note here the similarity between Fig. \ref{egge_temperature} (b) and the case of
 zero-temperature driving Fig. \ref{diagram_egge} (although differences in the values of $t_{ESD}$ and in the period of the lobes do exist). This can be qualitatively understood by using the fact that driving the system is mathematically equivalent with a combination of temperature, dephasing, and squeezing \cite{driven}, and therefore, for certain initial states, the driving could lead to similar effect as temperature.

 The reason for the appearance of the lobs is then similar to that given in Section \ref{eggesection}: consider for example $p=0$, which corresponds to the case in which one qubit
 in the excited state is connected to the other qubit lying in the ground state. What happens is that the single quanta
 of energy start to oscillate between the two qubits. The dipole-dipole form of the interaction assures that the dynamics occurs only in the subspace spanned by $|01\rangle$ and $|10\rangle$, and, during its evolution, the state of the system goes through Bell entangled states of the type $|\Psi^{\pm}\rangle = (|01\rangle \pm |10\rangle)/\sqrt{2}$. Now, if there is no dissipation, the concurrence will simply oscillate. If the temperature is zero, the concurrence decays exponentially. Once temperature is introduced, the system will display sudden death of entanglement and will acquire a finite $t_{ESD}$. Moreover, delayed sudden birth of entanglement is produced due to the fact that during oscillations the concurrence could go below zero but then revive after some time,  winning against the thermal noise.

\section{Conclusion}

We have investigated the phenomenon of sudden death of entanglement for certain subclasses of so-called X states under continuous driving and coupling. We show that
each of these states present notable peculiarities when analyzing the time evolution: for Werner states, the evolution is independent of coupling, for YE states the time of sudden death presents soft oscillations as a result of  coupling, and  for diagonal ($eg-ge$ and $ee-gg$) states $t_{ESD}$ shows abrupt oscillations, due to the sudden appearance of delayed birth of entanglement  for certain value of the coupling. We present both analytical formulae and numerical evidence for these phenomena.

\section{Acknowledgments}
This work was supported by the Academy of Finland (Acad. Res. Fellowship 00857 and projects 7111994 and 7118122).


\begin{thebibliography}{99}


\bibitem{book} R. R. Puri, {\em Mathematical Methods of Quantum Optics} (Springer,
Berlin, 2001).

\bibitem{Yu1}
T. Yu and J. H. Eberly, Phys. Rev. Lett. {\bf 93}, 140404 (2004).

\bibitem{Yu3}
T. Yu and J. H. Eberly, Phys. Rev. Lett. {\bf 97}, 140403 (2006).

\bibitem{Yu2}
T. Yu and J. H. Eberly, Quant. Inf. and Comp. {\bf 7}, 459 (2007).



\bibitem{almeida} M. P. Almeida, F. de Melo, M. Hor-Meyll, A. Salles, S. P. Walborn, P. H. Souto Ribeiro, L. Davidovich, Science {\bf 316}, 579 (2007) ;
J. H. Eberly and T. Yu, Science {\bf 316}, 555 (2007);
J.~Laurat, K.~S.~Choi, H.~Deng, C.~W.~Chou, and H.~J.~Kimble, Phys.~Rev.~Lett.~{\bf 99} 180504 (2007).

\bibitem{finitetemp} A. Al-Qasimi and D. F. V. James, Phys. Rev. A {\bf 77}, 012117 (2008).

\bibitem{driven} J. Li, K. Chalapat, and G. S. Paraoanu, J. Low Temp. Phys. {\bf 153}, 294 (2008).

\bibitem{direct} T.~Yamamoto, Yu.~A.~Paskin,
 O.~Astafiev, Y.~Nakamura, and J.~S.~Tsai, Nature {\bf 425}, 941 (2003); Yu.~A.~Paskin, T.~Yamamoto,
 O.~Astafiev, Y.~Nakamura, D.~V.~Averin, and J.~S.~Tsai, Nature {\bf 421}, 823 (2003); C. Rigetti, A. Blais, and M. Devoret, Phys. Rev. Lett. {\bf 94}, 240502 (2005);
G. S. Paraoanu, Phys. Rev. B {\bf 74}, 140504(R) (2006); J. Li, K. Chalapat, and  G.S. Paraoanu, Phys. Rev. B {\bf 78}, 064503 (2008).

\bibitem{indirect} A. Blais, J. Gambetta, A. Wallraff, D. I. Schuster, S. M. Girvin, M. H. Devoret, and R. J. Schoelkopf, Phys. Rev. A {\bf 75}, 032329 (2007);
J. Majer, J. M. Chow, J. M. Gambetta, Jens Koch, B. R. Johnson, J. A. Schreier,
L. Frunzio, D. I. Schuster, A. A. Houck, A. Wallraff, A. Blais, M. H. Devoret,
S. M. Girvin, R. J. Schoelkopf, Nature {\bf 449}, 443 (2007);
A. O. Niskanen, K. Harabi, F. Yoshihara, Y. Nakamura, S. Lloyd, and J. S. Tsai,
Science {\bf 316}, 723 (2007);


\bibitem{steadystate} N. Lambert, R. Aguado, and T. Brandes, Phys. Rev. B {\bf 75}, 045340 (2007); L. D. Contreras-Pulido and R. Aguado, Phys. Rev. B {\bf 77}, 155420 (2008); L. Hartmann, W. D\"urr, and H.-J. Briegel, Phys. Rev. A {\bf 74}, 052304 (2006); L. Hartmann, W. D\"urr, and H.-J. Briegel, New J. Phys. {\bf 9} 230 (2007); M. B. Plenio and S. F. Huelga, Phys. Rev. Lett. {\bf 88}, 197901 (2002); S. Huelga and M. Plenio, Phys.Rev. Lett. {\bf 98}, 170601 (2007);
    S. Mancini, S. Bose, Phys. Rev. A {\bf 70}, 022307 (2004); S. Mancini, and J. Wang, Eur. Phys. J. D {\bf 32} (2005);
D. G. Angelakis, S. Bose, and S. Mancini, Europhys. Lett. {\bf 85}, 20007 (2009); J. Li and G.S. Paraoanu, arXiv:0903.3464.









\bibitem{Cohen}
C. Cohen-Tannoudji, J. Dupont-Roc, and G. Grynberg, {\em Atom-Photon Interactions} (John Wiley, 1992).





\bibitem{werner} R. F. Werner, Phys. Rev. A {\bf 40}, 4277 (1989).









\bibitem{brassard}R. F. Werner, Phys. Rev. A {\bf 40}, 4277 (1989); C.~H.~Bennett {\it et al.}, Phys. Rev. Lett. {\bf 76}, 722 (1996).


\bibitem{Wootters}
W. K. Wootters, Phys. Rev. Lett. {\bf 80}, 2245 (1998).

\bibitem{agarwal} S. Das and G. S. Agarwal, arXiv:0901.2114






\end{thebibliography}
\end{document}